\documentclass[journal]{IEEEtran}
\usepackage{epsfig,epsf,rotating,setspace,latexsym,amsmath,amssymb,amsfonts,bm,theorem,subfigure,epstopdf}
\usepackage{cite,authblk}
\usepackage{bbm}
\usepackage{algorithm,statmath}
\usepackage[noend]{algpseudocode}
\usepackage{color}
\usepackage{mathtools}
\usepackage{soul}

\algrenewcommand\algorithmicforall{\textbf{foreach}}
\algrenewcommand\algorithmicindent{.8em}

\IEEEoverridecommandlockouts
\allowdisplaybreaks

\begin{document}

\title{Timeliness: A New Design Metric and \\ a New Attack Surface}

\author{Priyanka Kaswan \qquad Sennur Ulukus\\
        \normalsize Department of Electrical and Computer Engineering\\
        \normalsize University of Maryland, College Park, MD 20742\\
        \normalsize  \emph{pkaswan@umd.edu} \qquad
        \emph{ulukus@umd.edu}}

\maketitle

\begin{abstract}
As the landscape of time-sensitive applications gains prominence in 5G/6G communications, timeliness of information updates at network nodes has become crucial, which is popularly quantified in the literature by the age of information metric. However, as we devise policies to improve age of information of our systems, we inadvertently introduce a new vulnerability for adversaries to exploit. In this article, we comprehensively discuss the diverse threats that age-based systems are vulnerable to. We begin with discussion on densely interconnected networks that employ gossiping between nodes to expedite dissemination of dynamic information in the network, and show how the age-based nature of gossiping renders these networks uniquely susceptible to threats such as timestomping attacks, jamming attacks, and the propagation of misinformation. Later, we survey adversarial works within simpler network settings, specifically in one-hop and two-hop configurations, and delve into adversarial robustness concerning challenges posed by jamming, timestomping, and issues related to privacy leakage. We conclude this article with future directions that aim to address challenges posed by more intelligent adversaries and robustness of networks to them. 
\end{abstract}

\section{Introduction}
Next generation wireless networks will be characterized by dense interconnected infrastructure supporting applications such as autonomous driving, blockchains, internet of things (IoT), augmented reality (AR), virtual reality (VR), and remote healthcare (RH) applications, that demand real-time interaction. In face of network resource limitations and increasingly dynamic data generated by various sources in these networks, it is imperative that all nodes within these networks have the latest possible updates about the source nodes at all times for seamless functioning of these networks. 

\begin{figure}
    \centering
    \includegraphics[scale = 0.4]{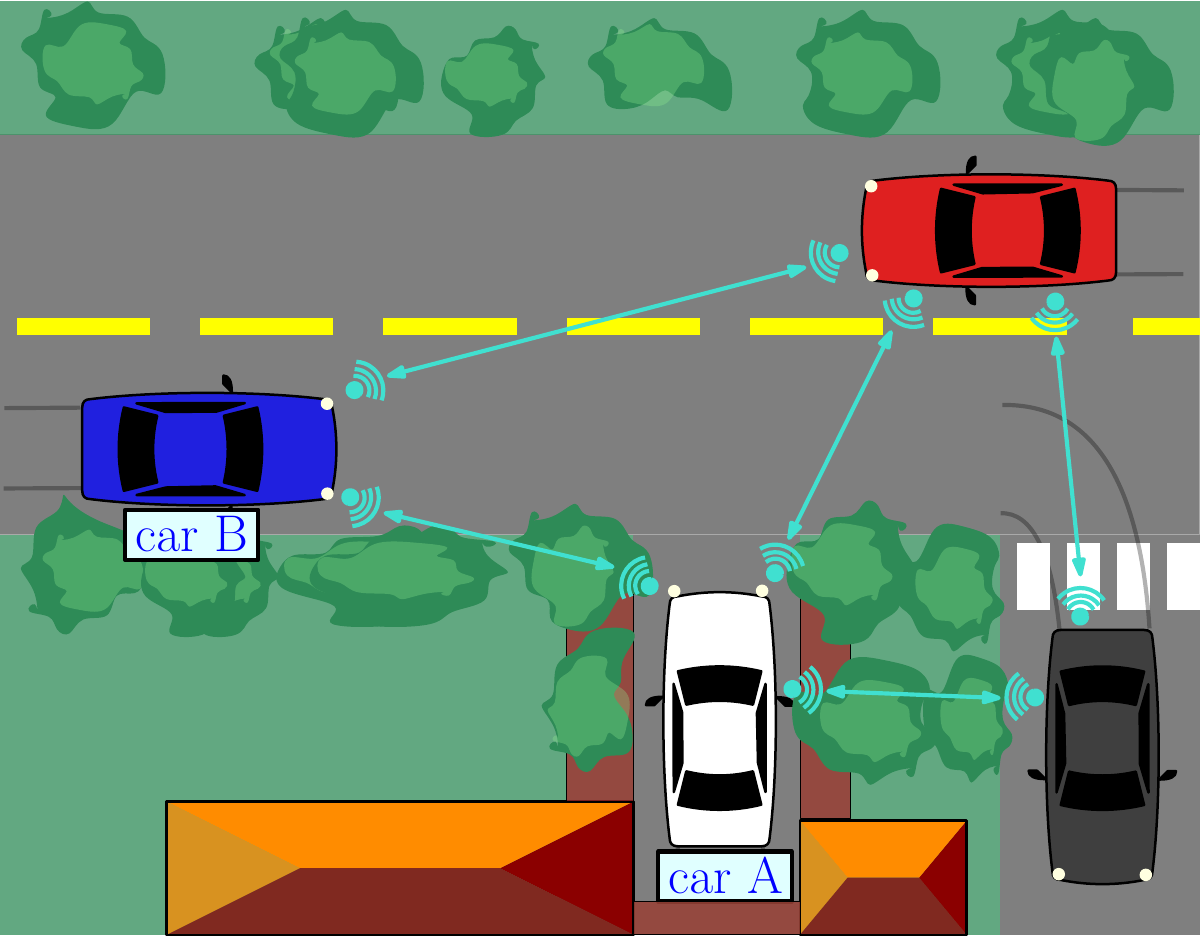}
    \caption{Timely communication is absolutely critical in autonomous driving.}
    \label{fig:neighbourhoodcars}
\end{figure}

Consider Fig.~\ref{fig:neighbourhoodcars}, where car B (blue car) approaches car A (white car) moving out of a driveway. To prevent a potential collision between the two cars, car A needs to be continuously aware of the precise real-time location of car B. However, the continuous change in car B's coordinates would necessitate an infinite number of instantaneous updates from car B to car A, which may not be possible due to several issues, bandwidth limitations being a prominent one. Even if car B attempts to transmit a high volume of updates rapidly (high throughput), these updates can form a backlog in the car radio's queue, hence by the time a queued update is serviced and transmitted, it becomes outdated. To minimize the waiting time of packets in the queue (low delay), one approach is to send packets infrequently, thereby keeping the queue mostly empty for almost instant reception of any packet sent by car B to car A. However, the infrequent updates result in car A maintaining an outdated view of car B's location for extended periods. Clearly, traditional metrics such as throughput and delay struggle to accurately capture the semantic property of the timeliness of updates. Therefore, \cite{kaul2012real} introduced a new metric - age of information (AoI), to assess the freshness of updates at a receiver node. If a node at time $t$ has a packet that was generated at the source at time $g(t)$, the instantaneous age of information at the node is defined as $a(t)=t-g(t)$. A more recent generation time implies a lower age, hence upon receiving a packet, a node compares its timestamp with the previously held packet, retaining only the fresher one and discarding the staler packet. 

However, a new design metric unlocks opportunities for new malfunctions and new attacks by adversaries that aim to disrupt the efficient functioning of these systems, by introducing staleness in the network. For example, the success of military operations depends on timely communication with all nodes of the defense network including unmanned vehicles, autonomous vehicles, AR supported wearables and devices, and weapon systems. Here, an adversary could jam the link between nodes, thereby preventing communication and engagement between those nodes. Worse, since packets are prioritized on the basis of their age or generation timestamps, by manipulating the timestamps of packets in and out of a remote AR device, the adversary can disrupt situational awareness and create confusion, hindering mission success. Further, additional errors may infiltrate an age-based system, such as packets getting mutated during inter-node transmission leading to misinformation. Privacy is also at risk, as adversaries could eavesdrop on packets and their timestamps, potentially compromising sensitive information.

This article provides a concise overview of existing research papers on threats to age-based systems. We investigate attacks in two primary categories. We first cover threats to dense interconnected networks which use decentralized algorithms such as gossiping to speed dissemination of timely data in the network, and show how such threats uniquely capitalize on the timeliness and gossiping aspects of file exchange protocols. Thereafter, we delve into simpler specific topologies, such as a transmitter-receiver pair. We conclude by offering insights into potential future directions into adversarial attacks and robustness of our communication networks.

\begin{figure}
    \centering
    \includegraphics[scale = 0.4]{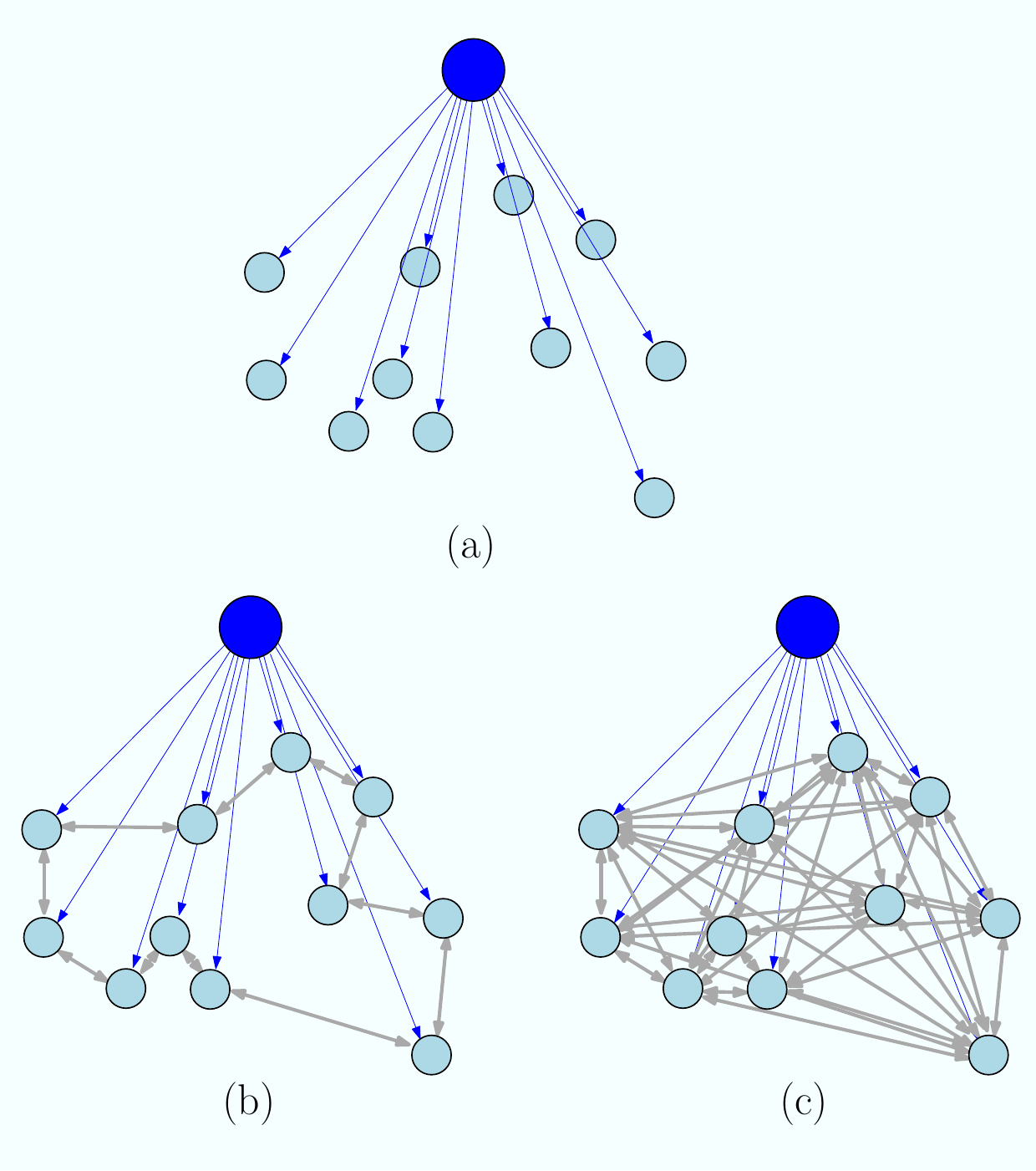}
    \caption{Expected age scales as (a) $O(n)$ in case of disconnected network, (b) $O(\sqrt{n})$ in case of ring network, (c) $O(\log{n})$ in case of fully connected network. The network consists of $n$ user nodes and a source node (dark blue).}
    \label{fig:3_gossip_networks}
\end{figure}

\section{Attacks on Dense Gossip Networks}
In large networks, it is often not feasible for a source node to send status updates to all nodes in a timely manner due to large network size and bandwidth limitation at the source. To overcome these limitations, simple scalable decentralized protocols such as gossip protocols have been proposed in the literature for expediting the dissemination of updates. In these protocols, each node recurrently chooses a node uniformly at random from its set of neighbors as a receiver to send its packet, thereby mimicking the spread of information like gossip or rumor. In age-based gossip protocols, the receiving node accepts a packet only if doing so results in a decrease in age at the receiver. We henceforth analyze the age of the network nodes with single type of files that are generated at a specific node, referred to as the source node. It has been shown in literature \cite{Yates21gossip_traditional} that in a network of $n$ nodes that wish to track a time-varying information source as in Fig.~\ref{fig:3_gossip_networks}, the age at each node scales as $O(n)$ if the network nodes are disconnected from one another (Fig.~\ref{fig:3_gossip_networks}(a)), since a larger $n$ implies that more nodes are dependent for updates on the source which has limited update capacity. However, if the nodes are connected to their neighboring nodes with whom they frequently gossip, then age at nodes can be reduced. Specifically, age scales as $O(\sqrt{n})$ and $O(\log{n})$ in a ring (Fig.~\ref{fig:3_gossip_networks}(b)) and fully-connected network (Fig.~\ref{fig:3_gossip_networks}(c)) topology, respectively. We next discuss threats that timely gossip networks face and how gossiping both acts as a shield against threats but also enables adversaries to propagate attacks more effectively. 

\subsection{Timestomping Attacks}
Timestomping corresponds to manipulation of timestamp of a packet, a term originating in the context of malware attacks where attackers alter the timestamp of a harmful file to evade detection during investigations. Timestomping can be uniquely injurious to age-based systems, since a receiver node compares the timestamp of every incoming packet with the timestamp of the packet already in its possession to only keep the packet with the fresher timestamp. Naturally, if the timestamps marked on these packets are manipulated, the receiver node can be tricked into picking a staler packet, thereby increasing its age. 

\begin{figure}
    \centering
    \includegraphics[scale = 0.4]{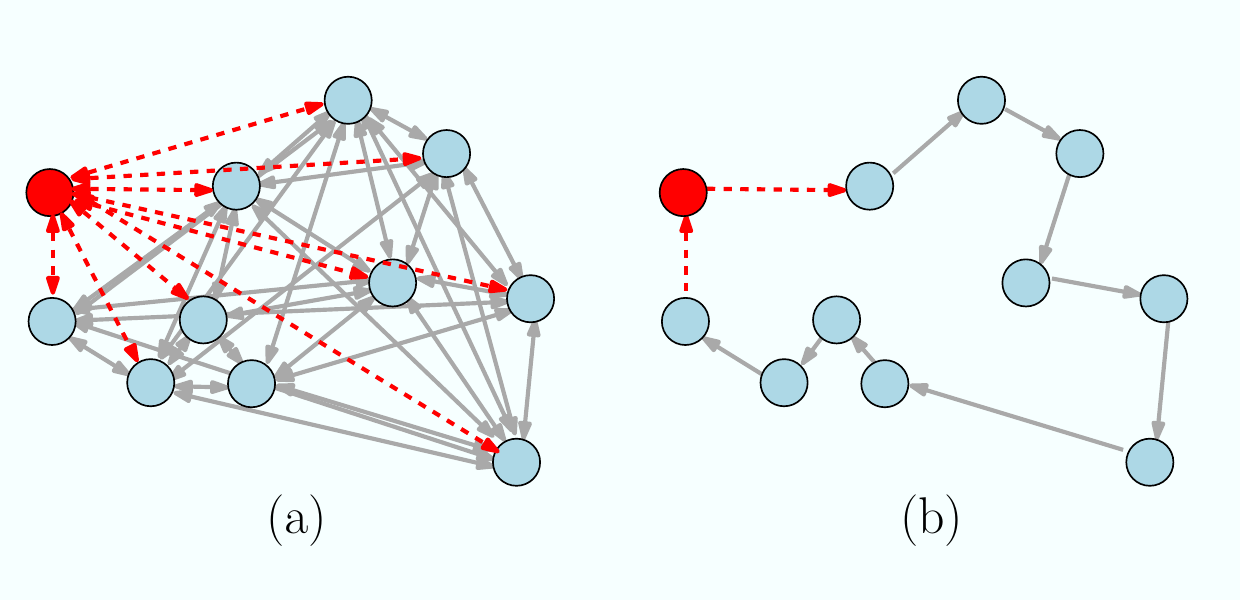}
    \caption{Red node captured by timestomping adversary in (a) fully connected gossip network, (b) unidirectional gossip network.}
    \label{fig:timestompingreview}
\end{figure}

The more the manipulated timestamp digresses away from the true timestamp, the higher is the likelihood of error in age-conformed packet selection across all nodes that the timestomped packet traverses through inter-node gossiping. At time $t$, the timestamp of a packet can utmost be manipulated to either $0$ (designating it as an outdated packet, making it unlikely to be chosen) or the current time $t$ (declaring it as the latest packet, ensuring its selection over any other packet). Building upon this idea, \cite{kaswan_timestomp_jrl} considers a gossip network with an adversary that captures one of the $n$ nodes (Fig.~\ref{fig:timestompingreview}) and probabilistically changes timestamps of every outgoing and incoming packet, according to $Bern(p)$ and $Bern(q)$, respectively, to either $0$ or $t$, and it is investigated how age scales at other network nodes as $p$ and $q$ vary.

This work reveals some intriguing insights about the potential of timestomping attacks through the use of a powerful analytical tool, stochastic hybrid system (SHS) framework, which helps analyze changes in the actual age and the marked timestamps of the packets possessed by nodes, as various inter-node transmissions occur. The actual age here depends on the true timestamp of a packet, representing the time it was generated at the source, which may differ from the timestamp marked on the packet due to timestomping. \cite{kaswan_timestomp_jrl} demonstrated how network topology enables an adversary to impact the network age. In a fully connected network (Fig.~\ref{fig:timestompingreview}(a)), it was shown how one such infected node can single-handedly suppress freshness of a large network of $n$ nodes, and increase the expected age from $O(\log n)$ (found in \cite{Yates21gossip_traditional}) to $O(n)$. The adversary's optimal strategy involves consistently increasing the timestamp of every outgoing packet to $t$ and decreasing the timestamp of every incoming packet to $0$. This effectively prevents the infected node from accepting any incoming files (thus, its information gets staler and staler) and encourages other nodes to always accept outgoing packets from the infected node (which are stale). However, if the infected node accepts even a small fraction of incoming packets, then a large network has the potential to suppress of the spread of stale files from the infected node by reducing its age to $O(\log{n})$. These findings show how full connectivity can be both beneficial and detrimental for the network under timestomping attacks. In a slightly modified network, \cite{kaswan_timestomp_jrl} considered an adversary which only intercepts and timestomps packets from the source to a specific network node, i.e., the adversary is not directly in contact with all $n$ nodes anymore. However, the network age still got deteriorated to $O(n)$, owing to the full connectivity in the network. This highlights how with little an effort an adversary can induce staleness in the whole network. 

On the other extreme of network connectivity, \cite{kaswan_timestomp_jrl} analyzes timestomping attack in a unidirectional ring (Fig.~\ref{fig:timestompingreview}(b)). Here the adversarial impact of the infected node is confined by its distance from the adversary, and regardless of adversarial policy, the age scaling for a larger fraction of the network continues to be $O(\sqrt{n})$, the same as age in a unidirectional ring without an adversary. Note that upon encountering a packet, the adversary in gossip network cannot tell whether the timestamp marked on the packet is true or manipulated through timestomping. Therefore, the timestomping adversary considered in this work is an oblivious adversary, i.e., it probabilistically changes timestamps without inspecting the packet timestamp beforehand. This contrasts with simpler networks explored later, where the adversary is fully aware of actual timestamps, allowing for more intelligent decision-making.

\subsection{Jamming Attacks}
\begin{figure}
    \centering
    \includegraphics[scale = 0.5]{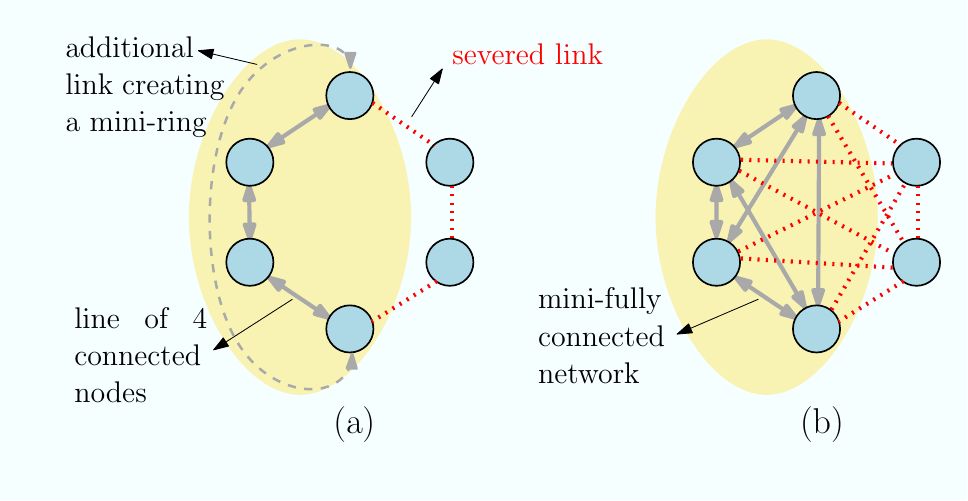}
    \caption{Jamming attack in (a) ring network, (b) fully connected network. The dotted red links represent the links jammed by adversary.}
    \label{fig:jammerreview}
\end{figure}

In jamming attacks, an adversary disrupts communication between two nodes by filling the channel with noise (though jammed links could also be a proxy for link failures). This causes a node to interact with a smaller set of neighbors, thereby altering the network topology. Hence, analyzing an age-based gossip network of given topology for jamming attacks is akin to analyzing a gossip network of modified topology with reduced inter-node connectivity. In this respect, \cite{kaswan_jamming_jrl} examines how the sum of age of all $n$ network nodes varies in presence of $\Tilde{n}$ jammers, each severing a unique inter-node gossip link (Fig.~\ref{fig:jammerreview}). The average age of a ring network without an adversary is known to scale as $O(\sqrt{n})$; this paper shows that in the presence of $O(\sqrt{n})$ jammers, the average age continues to be $O(\sqrt{n})$, i.e., the ring network is robust to $O(\sqrt{n})$ jammers. Further, in presence of $O(n^{\alpha})$ jammers with $\alpha\in(\frac{1}{2},1]$, the average age of ring scales as $O(n^{\alpha})$, irrespective of the specific positioning of jammers around the ring.

Proving these results involves establishing a series of structural results. When jammers cut links on the ring, the ring is dismembered into a collection of line networks, losing the circular symmetry of a full ring. Hence, characterizing total age of the network requires age analysis of these disconnected line networks. Analysis reveals that in line networks, age is lowest at the nodes towards the center and increases towards the peripheral nodes, a fact leveraged to derive lower and upper bounds on total age of line networks. These bounds are constant multiples of total age of a mini-ring network, which is formed by introducing an additional link between the peripheral nodes as in Fig.~\ref{fig:jammerreview}(a) to re-introduce circular symmetry in the line network. The circular symmetry of these mini-rings conveniently provides a tractable expression for the total age of the involved nodes. Consequently, the age of the dismembered ring is bounded by constant multiples of an alternate mini-ring representation of the dismembered ring. \cite{kaswan_jamming_jrl} suggests that to cause maximum damage to system age, jammers should cut links in a manner that consolidates the remaining links into one long line network and leaves some isolated nodes. On the other hand, the least damaging jammer positioning is the equidistant positioning of jammers around the ring. 

\cite{kaswan_jamming_jrl} later also observes the consolidation property in fully connected networks, where for maximum damage to system age, jammers should cut links in a manner that consolidates the remaining links into one large ball of a mini-fully connected network and leaves some isolated nodes, Fig.~\ref{fig:jammerreview}(b). Building on this notion, through a greedy strategy of jammer placement, the paper proves that the fully connected network is robust upto $O(n\log{n})$ jammers, as opposed to $O(\sqrt{n})$ jammers in ring. This can be attributed to the fact that fully-connected network has $\binom{n}{2}=\frac{n(n-1)}{2}$ links of less frequent traffic compared to $n$ links of higher traffic in ring, hence more jammers are required to block a similar traffic load in a fully connected network. Therefore, as opposed to timestomping attacks, where full connectivity enables an adversary to propagate attack effectively, in jamming attacks, full connectivity constraints the effectiveness of jammers.

\subsection{Information Mutation and Misinformation}
We saw how age-based gossip networks face unique threats like timestomping that capitalize on the file exchange protocol being dependent on timestamps or perceived age of the packets. Likewise, the age-specific nature of file exchange protocols also makes these networks susceptible to the propagation of misinformation. \cite{kaswan23mutation} focuses on a fully connected network (Fig.~\ref{fig:3_gossip_networks}(c)) where a packet is vulnerable to corruption during its propagation through gossiping. While information directly obtained from the source is assumed accurate, when a user node transmits its packet to another, the packet information mutates with probability $p$ into some misleading version of information. 

The file exchange protocol relies primarily on version age, where the receiving node decides to keep or discard a packet based on a comparison of version numbers. If the version numbers differ, the more recent version is chosen; when two packets have the same version number, the protocol prioritizes truth over misinformation. Thus, if either packet contains accurate information, it takes precedence. The key question is, what fraction of the network nodes on average would have accurate information, denoted by the quantity $F$, which serves as a metric for how widespread is misinformation in the network. The problem is again modelled using an SHS framework, wherein changes in accuracy and version number of the packet present at all network nodes are tracked as inter-node transitions occur. 

Some intriguing observations emerge from the analytical and numerical analysis of this model. Firstly, a large network size $n$ leads to more misinformation in the network. This may be attributed to the fact that in a large network, nodes receive less frequent accurate updates from the source, owing to limited update capacity of the source. In such scenarios, nodes heavily rely on packets arriving through the gossip mechanism, which is prone to information corruption. 

Next, if the source gets rapidly updated with newer versions, it results in higher misinformation spread in the network. Conversely, if the source rarely gets updated, then a single version prevails at the source for extended time-periods. In the latter case, since the source continues to disseminate accurate copies of its current version and since truth prevails over misinformation, any mutated packet at a node eventually gets replaced with the accurate copy of this version (received either from another node or the source directly). Hence, infrequent source version updates prevents misinformation. 

A high mutation probability $p$ understandably leads to higher misinformation spread. However, the extreme value of $p=1$ does not guarantee that all nodes will be misinformed. This is because the source consistently disseminates accurate information to the network nodes, which cannot be replaced by misinformation corresponding to the same or older versions, as truth prevails over misinformation.

The rate with which the source transmits packets to the network has a rather unintuitive impact on misinformation spread. If source transmits packets with high rate, then all network nodes will have the accurate current version prevailing at the source, which prevents misinformation spread since accurate information cannot be replaced by misinformation. However, very low source transmission rate also curbs misinformation. This is because if the source does not transmit a newer version to any network node for a long time, the last (accurate) update from the source to any network node marks the freshest packet in the network, which will slowly replace any inaccurate or staler information in the network, thereby curbing misinformation. Thus, somewhat counterintuitively, misinformation spread is higher for moderate source transmission rates.

Perhaps the highlight of \cite{kaswan23mutation}  is how gossiping enables or curbs spread of misinformation. On one hand, small gossiping rates nullify the chances of generation of mutated information, and the network in this case primarily relies on the fresh accurate packets from the source; this aligns with how this threat is uniquely applicable to age-based gossip networks. Interestingly, very high gossip rates also prove effective against misinformation, since any new version from the source to any node is instantly circulated to all network nodes, replacing any misleading or stale information. Thus, just as in the case of source transmission rates, misinformation spread is higher for moderate inter-node gossiping rates.

\section{Threats in Simpler Age-Based Systems}
We now survey adversarial works for simpler networks, such as source-transmitter pair(s), with occasional presence of relaying cache node. The leading work in this regard is \cite{nguyen2017hostile_inf} which explored the impact of hostile interference on age at a receiving node. The system model consists of a transmitter, a receiver and an interferer, and a two-player game is considered, such that one player—the transmitter—strives to maintain the freshness of information at the receiver by sending updates, while the other player—the interferer—aims to degrade the freshness of information at the receiver.

The strategy of each player is the power level transmitted by the player: the transmitter's strategy is a transmission power level $p$, and the interferer's strategy is an interfering power level $q$. The receiver is affected by hostile interference from the interferer and a noise of power $N$, such that the transmitter is able to transmit with a bit rate that is an increasing function of the resulting SINR. Depending on the length of the packets, this limitation on bit rate puts a limitation on transmission rate of packets $\mu=\mu(p,q)$. The update packets are assumed to be generated at the transmitter according to a Poisson process with rate $\lambda$, therefore transmission of packets from the transmitter to the receiver is modelled as an $M/G/1/1$ queue, with packet arrival rate $\lambda$ and packet service rate $\mu$. The corresponding average peak age turns out to be $A(p,q)=\frac{1}{\lambda} + \frac{2}{\mu}$. The transmitter wants to reduce $A(p,q)$ while the interferer wants to increase $A(p,q)$. Each player however incurs a cost proportional to the power used by that player. Therefore, players choose a transmission power level that optimizes their respective utility function, which incorporates both the age-based reward and the operational cost. 

The paper aims to determine equilibria in two forms of pure strategies: Nash equilibrium (NE) and Stackelberg equilibrium (SE). Under simultaneous actions, NE is the point where neither player can enhance performance by altering the strategy while the other player's strategy remains constant. Alternatively, in a sequential setting, where one player acts as the leader and chooses its strategy first, and the other player, the follower, reacts accordingly, the solution is the SE. \cite{nguyen2017hostile_inf} showed that when receiver noise power $N$ is zero, the NE exists and is unique, such that the power cost is the same for both transmitter and receiver at NE. On the other hand, SE  does not exist when transmiter is the leader, but exists when interferer is the leader, and the utility function of the latter SE exceeds the NE utility function. Both NE and SE do not depend on packet arrival rate $\lambda$. \cite{garnaev2019jamming_yates} later on extends this model by considering non-zero background noise and a slightly modified utility function, where it is shown that NE also depends on transmitter packet updating rate $\lambda$. Further, it proves that in contrast to NE, multiple SE arise in this model.

In contrast to \cite{nguyen2017hostile_inf} and \cite{garnaev2019jamming_yates} which consider a static game model, \cite{xiao2018dynamic_game} considers a dynamic jamming game between an attacker and the transmitter. In \cite{xiao2018dynamic_game}, the transmitter samples a physical process and sends the samples to timely update the receiver, while the attacker aims to sabotage the timely updates by jamming the channel. The attacker monitors the channel and detects every transmission, however the samples contain a small amount of information causing the duration of their transmission to be negligible, which makes it difficult for the attacker to target the transmission. Hence, after detecting the transmission of the $n$th sample, the attacker occupies the channel for some random duration of time $A_n$ in order to delay delivery of the next sample. The attacker's jamming times $A_n$ are assumed to be i.i.d.~following a density function $f_A$, and are limited by constraints of maximum and average jamming time (elusivity and energy constraints). The transmitter employs carrier sensing to monitor the channel, so once the channel is idle (i.e., after the attacker stops occupying the channel), the transmitter samples the physical process after a delay time of $D_n$ and transmits in the $n$th stage. The dynamic game between the system and the attacker consists of an infinite number of stages, where in each stage, the attacker takes action by choosing a jamming time distribution $f_A$ and the system reacts by choosing delay $D_n$, i.e., each stage game is a Stackelberg game with attacker taking action first. Under the equilibrium policy, it is shown that if the attacker's jamming time $A_n$ is short, the system chooses a larger delay $D_n$, conversely, the system chooses a smaller delay if the attacker's jamming time is long. Essentially, the transmitter sampling policy aims to equalize time differences between subsequent update deliveries. The attacker seeks to hinder the transmitter's ability to achieve this. Hence, the worst jamming distribution for the attacker would be fixed deterministic jamming time, making it easy for the transmitter to ensure periodic updates, and the best jamming distribution for the attacker would have high variance.

\begin{figure}
    \centering
    \includegraphics[scale = 0.6]{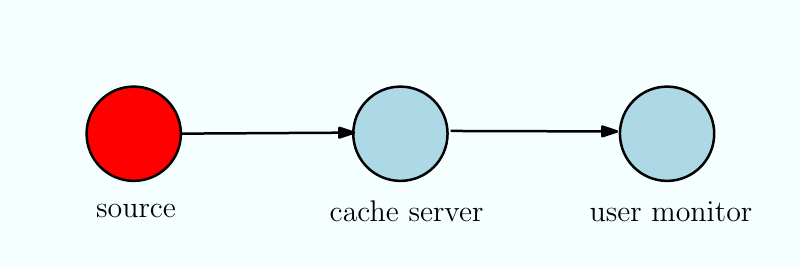}
    \caption{A $2$-hop network where information flows from source (red) to the end user through an intermediate server.}
    \label{fig:2hop_privacy}
\end{figure}

The works so far in this section analyzed jamming attacks in a system model with a single transmitter-receiver pair. In \cite{Banerjee22adversary}, the focus shifts to the system model of Fig.~\ref{fig:3_gossip_networks}(a), where one base station (BS) serves $N$ users over a wireless channel, in the presence of a jamming adversary. Each user has a dedicated channel and in each timeslot, the BS can communicate with only one user. The adversary can block at most one user in each timeslot and can only block upto $\alpha$ fraction of total $T$ slots (the time horizon for communication) due to power constraints. The oblivious adversary aims to increase the average age at users through an appropriate jamming sequence and the scheduling algorithm of BS aims to reduce the age. \cite{Banerjee22adversary} derives lower bounds on average age for deterministic and randomized user scheduling algorithms. Further, for randomized algorithms, it is shown that optimum action for the adversary is to block one specific user for consecutive $\alpha T$ slots at the center of the time horizon $T$. 

The paper also considers a modified system model where BS has $N_{sub}$ sub-carriers, such that in each timeslot, BS chooses one of those sub-carriers to send an update to one of the users. The adversary can block one of the $N_{sub}$ communication channels and can block upto $\alpha T$ timeslots. It is shown that the adversary should block $\alpha T$ slots of a specific sub-carrier in this case. The paper designs efficient online policies for both the models with provable performance guarantees. \cite{Banerjee22adversary_game} extends the system models of \cite{Banerjee22adversary} to a game-theoretic setting, where NE and SE are identified when they exist. With a setup similar to \cite{Banerjee22adversary} and \cite{Banerjee22adversary_game}, the BS in \cite{Banerjee22power_adversary} has $n$ discrete transmission power levels to communicate to the users, and the adversary has $m$ discrete blocking power levels to block the communication channels, such that probability of successful transmission of a packet depends on these power levels. Additionally, \cite{Banerjee20fundamentallimits,bhattacharjee2020competitive-short} competitively optimize age in cellular wireless networks with $N$ user equipments (UEs) under coverage of $M$ base stations (BS), such that an adversary decides whether a user channel will be good or bad in each timeslot in an online manner. In these works, an age optimization problem is considered and near-optimal competitive scheduling policies are designed.

Apart from jamming attacks, \cite{Banerjee22timestomping} explored timestomping attacks on a simplified communication model of Fig.~\ref{fig:2hop_privacy}, where a source attempts to minimize the age of a user. Due to a power constraint, the source can only transmit updates directly to the user for just $T_1$ timeslots over a time horizon of $T$. A cache node, which can afford more frequent transmissions, lies in between the source and the user, however the communication link between the cache and the user is under attack by a timestomping adversary. This adversarial cache updating problem is formulated as an online learning problem and competitive ratio for the problem is studied. Unlike the timestomping works in gossip network contexts, in this work, the adversary has full knowledge of true timestamps at the user and the packet transmission policy at the source, which allows the adversary to choose its timestomping action in every discrete timeslot more intelligently. 

On a different note, \cite{yates22privacy} sheds light on the privacy concerns associated with sending timely updates. \cite{yates22privacy} considers a system model similar to Fig.~\ref{fig:2hop_privacy}, where a source generates time-stamped updates that are sent to a server and then forwarded to a monitor. In the presence of an adversary capable of inferring information about the source by observing the server's output process, the server aims to release updates to a receiver while minimizing the information leaked to the adversary. This paper investigates the trade-offs between privacy and age for specific server policies.

\section{Conclusion}
Development of age-optimal policies is crucial for efficient operation of time-sensitive systems. However, we saw how the very mechanisms that make these networks efficient can become vulnerabilities under certain attacks. We surveyed works regarding threats to large gossip networks in context of jamming, timestomping and information mutation. We saw how jamming attacks are less effective against high-connectivity network topologies, but timestomping attacks can take advantage of high network connectivity to induce more staleness in the network. We also saw how age-based gossiping peculiarly impacts misinformation spread, as both very high or very low gossiping rates curb spread of misinformation and moderate gossiping rates aggrandizes misinformation. We then identified adversarial works on simpler age-based systems, investigating threats such as hostile interference, jamming attacks, timestomping attacks and privacy leakage.

\section{Future Directions}
In age-based systems where packet replacement relies on freshness of timestamps, timestomping attacks pose a significant threat. Current timestomping works manipulate the timestamps to either current time $t$ or outdated timestamp $0$, and the adversary takes the timestomping action probabilistically in gossip networks without looking into the packet. In place of such an oblivious adversary, there can be a smart adversary, which intelligently estimates the true timestamps of packets it encounters to decide whether a timestamp should be increased or reduced. Further, manipulating timestamps to $t$ makes it easy to detect the malfunctioning node, as receiving a packet with timestamp $t$ from a non-source node is a low-probability event. Hence, age characterization in the case of timestamp manipulation to more intermediate time instances is needed. Further, existing jamming studies analyze network age with static jammers, and an interesting future direction would be to see whether system age worsens if jammers dynamically switch to different links instead of sticking with one link. Further, though some specific policies have been analyzed from age-leakage perspective, we need to find the optimal policies to navigate the trade-off between privacy and age of information, especially in large gossip networks. 

\bibliographystyle{unsrt}
\bibliography{references}

\end{document}